\documentclass[apjl]{emulateapj}
\usepackage{graphicx}
\usepackage{subfigure}
\usepackage{amsmath, amssymb}
\usepackage{natbib}

\shorttitle{Tidal Disruption Rates in an E+A Galaxy}
\shortauthors{Stone and van Velzen}

\begin{document}

\title[Tidal Disruption Rates in an E+A Galaxy]
{An enhanced rate of tidal disruptions in the centrally overdense E+A galaxy NGC~3156}

\author{Nicholas C. Stone\altaffilmark{1}}
\affil{Columbia Astrophysics Laboratory, Columbia University, New York, NY, 10027, USA}

\and

\author{Sjoert van Velzen\altaffilmark{2}}
\affil{Department of Physics and Astronomy, The Johns Hopkins University, Baltimore, MD, 21218, USA}

\altaffiltext{1}{Einstein Fellow; nstone@phys.columbia.edu}
\altaffiltext{2}{Hubble Fellow; sjoert@jhu.edu}

\begin{abstract}
Time domain optical surveys have discovered roughly a dozen candidate stellar tidal disruption flares in the last five years, and future surveys like the {\it Large Synoptic Survey Telescope} will likely find hundreds to thousands more.  These tidal disruption events (TDEs) present an interesting puzzle: a majority of the current TDE sample is hosted by rare post-starburst galaxies, and tens of percent are hosted in even rarer E+A galaxies, which make up $\sim 0.1\%$ of all galaxies in the local universe.  E+As are therefore overrepresented among TDE hosts by 1-2 orders of magnitude, a discrepancy unlikely to be accounted for by selection effects.  We analyze {\it Hubble Space Telescope} photometry of one of the nearest E+A galaxies, NGC~3156, to estimate the rate of stellar tidal disruption produced as two-body relaxation diffuses stars onto orbits in the loss cone of the central supermassive black hole.  The rate of TDEs produced by two-body relaxation in NGC~3156 is large when compared to other galaxies with similar black hole mass: $\dot{N}_{\rm TDE}\sim 1\times 10^{-3}~{\rm yr}^{-1}$.  This suggests that the preference of TDEs for E+A hosts may be due to central stellar overdensities produced in recent starbursts.

\end{abstract}

\section{Introduction}
Stars are tidally disrupted in galactic nuclei when orbital perturbations reduce their angular momenta and place them on nearly radial orbits.  Once the stellar pericenter is reduced below a critical value, a strong tidal encounter with the central supermassive black hole (SMBH) destroys the star during its pericenter passage \citep{Hills75}.  Roughly half of the stellar mass falls back onto the SMBH, circularizing into an accretion disk and powering a luminous flare \citep{Rees88}.  In the last two decades, roughly a dozen candidate tidal disruption events (TDEs) have been discovered \citep{Bade+96, KomGre99, Esquej+07, Esquej+08, Maksym+10, Maksym+13} through the soft X-ray emission that is thought to be near the peak of their spectral energy distributions.  A comparable number have been found in the last decade via optical \citep{vanVel+11, Gezari+12, Chorno+14, Arcavi+14, Holoie+14, Holoie+16} or UV emission \citep{Gezari+06, Gezari+08} and upcoming time domain optical surveys are expected to discover hundreds to thousands more \citep{StrQua09, vanVel+11}.

Several dynamical processes are capable of feeding stars to SMBHs.  The most ubiquitous and robustly understood is two-body relaxation, which slowly diffuses stars through orbital phase space and eventually into the ``loss cone,'' the phase space region where stars can be ripped apart by tides from the SMBH \citep{FraRee76}.  Two-body relaxation calculations of TDE rates in realistic galaxies find that they are rare events, typically occurring roughly once per $10^{4-5}~{\rm yr}$ \citep{WanMer04, StoMet14}, with the highest rates in low-mass galaxies.  Observational estimates for the TDE rate are typically $\sim 10^{-5}~{\rm yr}^{-1}~{\rm galaxy}^{-1}$ \citep{Donley+02, vanFar14, KhaSaz14}, 
a number discrepant with theoretical estimates by an order of magnitude or more \citep{StoMet14}.  Other processes can in principle enhance the TDE rate above the floor set by two-body relaxation, such as non-conservation of angular momentum in axisymmetric or triaxial potentials \citep{MagTre99, MerPoo04}, interactions with massive perturbers such as molecular clouds \citep{Perets+07} and large-scale accretion disks \citep{KarSub07}, or gravitational wave recoil of the central SMBH \citep{StoLoe11}.  The impact of these more exotic mechanisms is more difficult to quantify observationally.

Recently, the sample of three TDE candidates discovered by \citet{Arcavi+14} using Palomar Transient Factory \citep{Law+09} data has identified an interesting puzzle: two of these TDEs are hosted by E+A galaxies, a relatively rare subtype of elliptical galaxy that has recently undergone a major starburst.  Although there is some dependence on the exact E+A definition used, these galaxies make up a fraction $10^{-2} > f_{\rm E+A} > 10^{-4}$ of all galaxies in the local universe \citep{Quinte+04}, so their overrepresentation in the \citet{Arcavi+14} sample indicates an elevated rate of tidal disruption.  Subsequent spectroscopic characterization of other TDE hosts found that {\it a majority} of all TDE flares inhabit ``Balmer-strong'' galaxies showing (i) no evidence of ongoing star formation but also (ii) Balmer absorption lines of unusually large equivalent width \citep{French+16}.  The absorption lines seen in Balmer-strong galaxies arise from a large population of A stars; the short lifetimes of these massive stars indicate that their host galaxies went through a major star formation episode $\sim 0.1-1~{\rm Gyr}$ in the past \citep{Snyder+11}.  A large minority ($3$ out of $8$) of the TDE hosts studied in \citet{French+16} are formally E+As, allowing an event rate of $\dot{N}\sim 10^{-3}~{\rm yr}^{-1}$ to be inferred for these galaxies, which is two orders of magnitude higher than the observed TDE rate for all types of galaxies.  The extreme overrepresentation of TDE candidates in rare galaxy subtypes worsens the pre-existing rate discrepancy for normal galaxies.

Several authors have speculated about dynamical mechanisms that could enhance the intrinsic rate of TDEs in post-starburst galaxies.  If a galaxy merger creates a SMBH binary in the center of the merger product, a short-lived ($10^{5-6}~{\rm yr}$) phase of greatly enhanced TDE rates will ensue, due partially to the Kozai effect \citep{Ivanov+05} but mostly to chaotic three-body orbits \citep{Chen+11}.  \citet{Arcavi+14} hypothesized that if a major merger triggers the starburst, E+As may overproduce TDEs due to the presence of hardening SMBH binaries.  However, although these binaries can enhance TDE rates up to $\dot{N}\sim 10^{-1}~{\rm yr}^{-1}$, the short duration of this enhancement means that SMBH binaries likely contribute only $\sim 1\%$ of the volumetric TDE rate \citep{WegBod11}.  Furthermore, it is unclear whether most SMBH binaries should exist in E+As; if the final parsec problem is solved very efficiently (inefficiently) then it is possible that most such binaries merge before (after) their host reaches the E+A stage.

Another possibility is that the starburst that created the E+A involved the dissipative flow of gas to the galactic nucleus, creating a steep stellar density cusp.  The denser the stellar population, the shorter the two-body relaxation time and the higher the TDE rate.  The starbursts that create E+As are quite substantial, increasing the stellar mass of the galaxy by $\sim 10\%$ \citep{Swinba+12}, so they are therefore quite capable of creating changes of order unity in the stellar density profile on parsec scales, where most TDEs are sourced.  Both multi-band photometry \citep{Yang+06} and resolved spectroscopy \citep{Pracy+12} of nearby E+As find significant radial gradients in stellar age, indicating an overabundance of young stars in E+A centers relative to their outskirts, and lending further plausibility to the idea of a central overdensity.  This hypothesis, first advanced by \citet{StoMet14}, can be tested by high-resolution photometric observations of the nearest E+A galaxies, and systematic calculation of TDE rates in their nuclei.

\begin{figure}
\includegraphics[width=85mm]{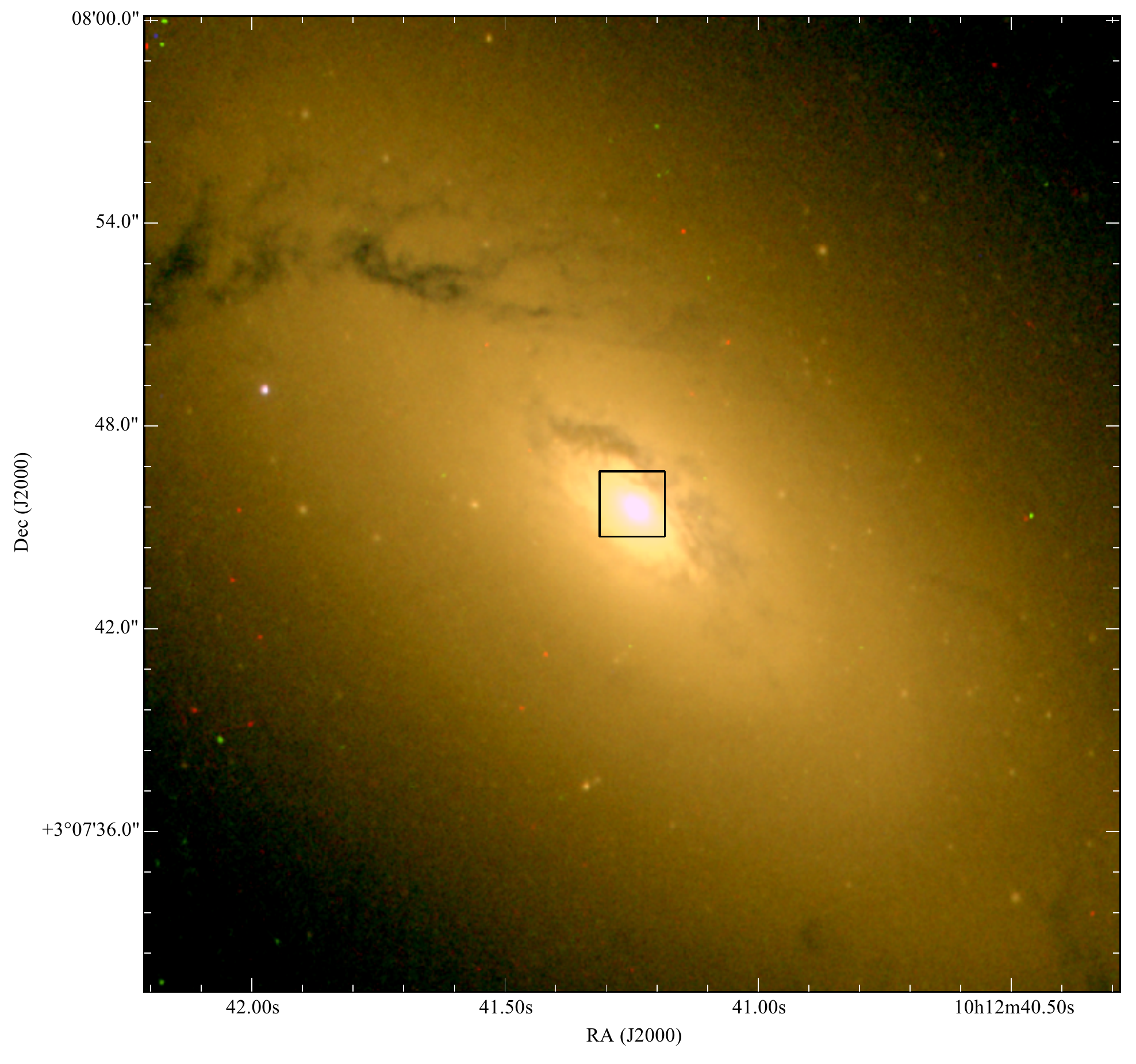}
\caption{A false-color image of NGC~3156, combining {\it HST} photometry in the F475W, F555W, and F814W filters.  The dimensions of the entire image are 1500~pc on each side; the smaller box in the middle is 200~pc on each side and is shown in Figures 2 and 3.}
\label{fig:NGC3156}
\end{figure}

Fortunately, one of the nearest E+A galaxies \citep{Pracy+12}, 
NGC~3156 (shown in Fig.~\ref{fig:NGC3156}), has been the target of past {\it Hubble Space Telescope (HST)} photometry ({\it HST} Program 12500; PI Kaviraj).  In this paper, we use archival {\it HST} data to estimate TDE rates in this galaxy, which at first glance appears to be an extreme outlier in terms of central stellar density.  We outline the {\it HST} observations, their uncertainties, and the range of allowable surface brightness profiles for this galaxy in \S \ref{sec:obs}.  In \S \ref{sec:rate}, we compute TDE rates in NGC~3156 across the range of allowable surface brightness profiles.  These rates are sensitive to the inward extrapolation of surface brightness (beyond the {\it HST} resolution limit), and we consider a range of theoretically motivated extrapolations.  In \S \ref{sec:discussion}, we discuss both the limitations and the broader implications of our analysis.

\section{NGC~3156: Observations}
\label{sec:obs}
Below, we first present an estimate of the mass of the central black hole in NGC~3156, followed by a discussion of the surface brightness profile inferred from {\it HST} observations of this galaxy.  {Finally, we note that NGC~3156 is a type~II Seyfert galaxy; the narrow [O\,\textsc{iii}] emission line  ($L_{[{\rm O\, \textsc{iii}}]}=2 \times 10^{38}\,{\rm erg}\,{\rm s}^{-1}$) is large compared to $H{\beta}$ emission (which is dominated by absorption), but no broad emission lines are observed, suggesting that our view of the accretion disk is obscured by dust.  We conclude this section by showing that unresolved optical emission from the central active galactic nucleus (AGN) is very small and can be neglected in our analysis of the surface brightness profile.}
\subsection{Black hole mass}
NGC~3156 has a $V$-band absolute magnitude of $M_{\rm V} = -19.4$ \citep{deVauco+91}.  Using the SDSS $u$,\,$g$,\,$r$,\,$i$,\,$z$ photometry \citep{Fukugi+96, York+00, Adelma+08} and the \verb kcorrect  software \citep{BlaRow07}, 
we estimate a galaxy-averaged mass-to-light ratio of $\Upsilon = 1.58$, in good agreement with Jeans and Schwarzschild modeling of this galaxy \citep{Cappel+06}.  This gives a stellar mass $M_{\rm tot}=3.6\times 10^{9}M_\odot$, which translates into an SMBH mass of $M_\bullet = 1.0\times 10^7 M_\odot$ if we associate this with the bulge mass $M_{\rm b}$ and use the $M_\bullet$-$M_{\rm b}$ scaling relation of \citet{KorHo13}, or a mass $M_\bullet = 8.9 \times 10^6 M_\odot$ if we instead use \citet{McCMa13}.  

However, observed E+A galaxies often possess a significant disk component \citep{Yang+04}, implying that the above estimate is likely an upper limit to $M_\bullet$.  To estimate the bulge-to-total ratio of the galaxy, we model the surface brightness profile with an exponential and a de Vaucouleurs profile (i.e., a S{\'e}rsic profile with $n=1$ and $n=4$). We find S{\'e}rsic radii ($R_e$) of 902 and 53.7 pc for the exponential and de Vaucouleurs profiles, respectively. The ratio of flux in the exponential and the de Vaucouleurs component is a factor of 7. Assuming that the integrated luminosity of the de Vaucouleurs profile provides a good description of the bulge mass, the implied black hole mass is $M_\bullet \approx 0.94 - 1.0 \times 10^6~M_\odot$, depending on the choice of calibration for the $M_\bullet$-$M_{\rm b}$ relation.  The disk-dominated nature of NGC 3156 has already been pointed out by \citet{Cappel+07}.  The disk component is unimportant for our analysis, since it presents a negligible contribution to the surface brightness in the inner 100~pc of the galaxy.
 
The $M_\bullet$-$\sigma$ relation offers an alternate avenue to estimate SMBH masses.  A central velocity dispersion of $\sigma=68~{\rm km~s}^{-1}$ was measured by \citet{Cappel+06} and \citet{Cappel+13}, which gives $M_\bullet = 2.7\times 10^6 M_\odot$ using the \citet{KorHo13} calibration of the $M_\bullet$-$\sigma$ relation, in reasonably good agreement with our application of the $M_\bullet$-$M_{\rm b}$ relation.  We take $M_\bullet =2.7\times 10^6 M_\odot$ as our fiducial value because of the greater uncertainties associated with a bulge-disk decomposition, but shall demonstrate that our results are not especially sensitive to this choice.

\begin{figure*}
\centering
\subfigure[ATLAS3D parametrization]{\includegraphics[width=85mm]{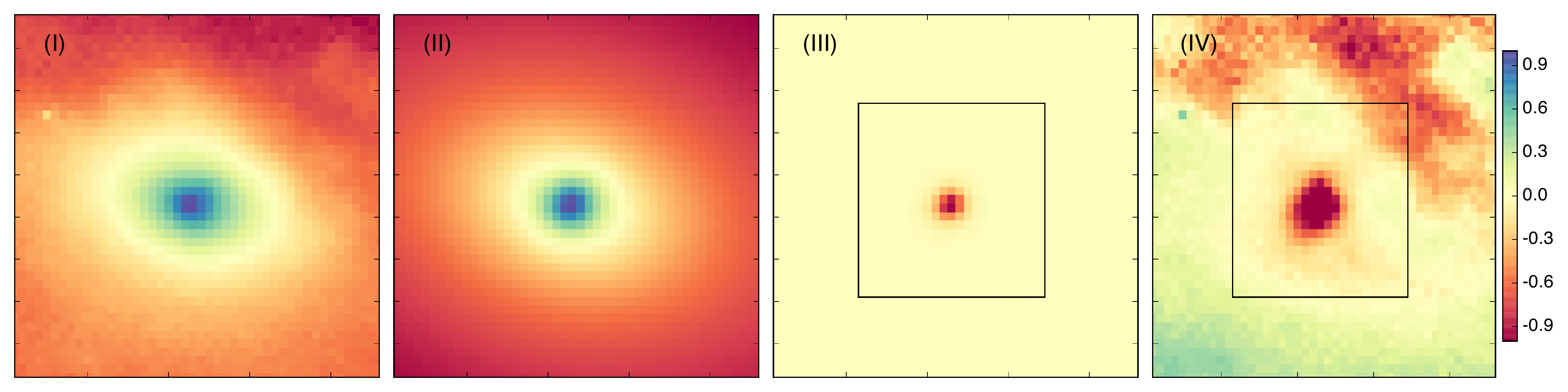}}\label{fig:NGC3156Atlas} \quad
\subfigure[This work]{\includegraphics[width=85mm]{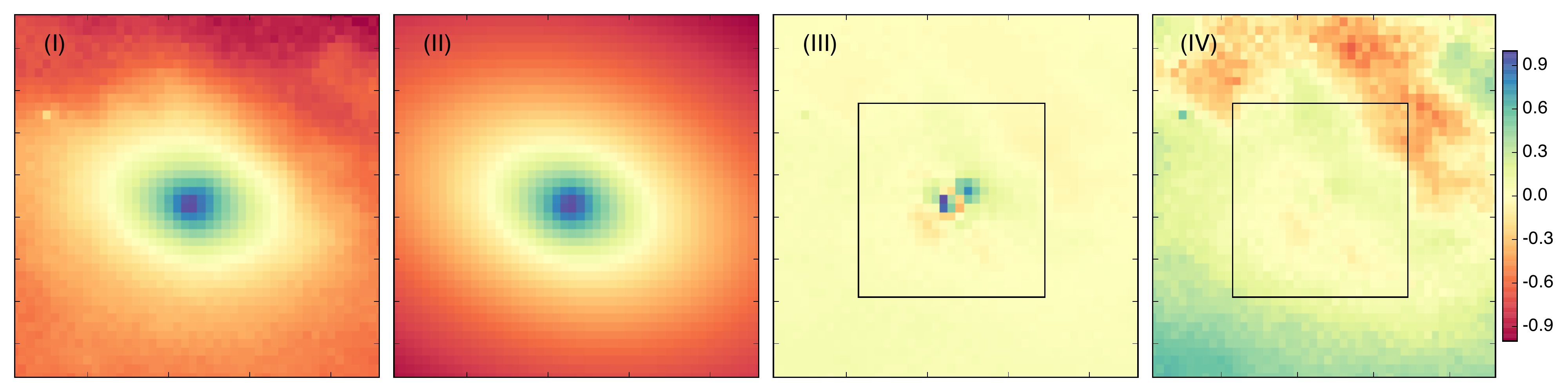}}\label{fig:NGC3156Sjoert}
\caption{The innermost 200~pc of NGC~3156. The box indicates the 100~pc region that is used in our fit for the parameters of the surface brightness profile.  Panels (I) and (II) show the WFC3/F475W observations and the model, respectively (on an arcsinh stretch). Panels (III) and (IV) show absolute and relative residuals for this fit, respectively (on a linear scale). The color scale of relative residuals ranges from -100\% to +100\%, as indicated by the bar. Large negative residuals are visible in the central pixels of the ATLAS3D surface brightness model, indicating that the fit severely overestimates the resolved and unresolved light from the very center of the galaxy. Our parametrization yields a flattened inner power law ($\gamma =1.2$) inside the break radius $R_{\rm b} = 20~{\rm pc}$, which  provides a much better parametrization of the observations, as can be seen by the relatively small residuals. A dust lane is visible in the top right of panels (I) and (IV).}
\end{figure*}

\subsection{Surface brightness profile derived from HST observations}
The E+A galaxy NGC 3156 was observed with {\it HST} WFC3 imaging in cycle 19, using the F225W, F475W, F555W, and F814W filters. 
The galaxy's surface brightness profile $I(R)$ was fit to the Nuker parametrization,
\begin{equation}\label{eq:nuker}
I(R) = 2^{(\beta-\gamma)/\alpha}I_{\rm b}\left(\frac{R_{\rm b}}{R} \right)^\gamma \left[1+\left(\frac{R}{R_{\rm b}} \right)^\alpha \right]^{(\gamma-\beta)/\alpha}, 
\end{equation}
by \citet{Krajno+13}, who found a projected break radius $R_{\rm b}=89.2~{\rm pc}$, a surface brightness at the break $I_{\rm b} = 2093L_\odot/~{\rm pc}^2$, an inner power-law slope of $\gamma = 1.78$, an outer power-law slope of $\beta = 0.86$, and a mediating power-law index (or softening parameter) of $\alpha = 4.29$.  Its observed redshift is $z=0.004463$, corresponding to a distance of 22~Mpc after correcting for peculiar velocities \citep{Blanto+05} and adopting a Hubble constant $H_{0}=70\,{\rm km}\,{\rm s}^{-1}{\rm Mpc}^{-1}$.

Notably, the inner power-law index $\gamma$ value marks NGC~3156 as an extreme outlier; the steepest central density cusp seen in the sample of $N=219$ galaxies analyzed by \citet{StoMet14} was $\gamma = 1.12$.  A naive inward extrapolation of the fiducial Nuker fit given here would predict that NGC~3156 is a TDE factory.  Indeed, the TDE rate diverges at small $R$ when $\gamma > 5/4$ \citep{SyeUlm99}, as is the case here.  
Fig. 2(a) shows the {\it HST} image of the inner 200~pc of NGC~3156, the fiducial Nuker fit of \citet{Krajno+13}, and the residuals of this fit.  The large negative residuals in the innermost pixels indicate that the fiducial fit is severely overestimating the innermost light (and therefore would also overestimate the TDE rate). 
To motivate a more careful analysis, we note that, empirically, TDEs are primarily sourced from a critical radius $r_{\rm crit}$, which is comparable to the SMBH influence radius\footnote{Defined in this paper as the radius containing a mass in stars equal to $M_\bullet$.} $r_{\rm infl}$.  The scaling relations of \citet{StoMet14} predict an influence radius $r_{\rm infl} \approx 3~{\rm pc}$ for NGC~3156, just under the WFC3 pixel size of 0.04'' ($4.3~{\rm pc}$).  Clearly, careful treatment of the innermost pixels is warranted.

\subsubsection{Point-Spread Function (PSF) models}
To measure the surface brightness profile of the innermost regions of the galaxy, we need an accurate representation of the PSF of the images. We identified one relatively bright star in the field of view (SDSS~J101237.70+030724.1, detected with a signal-to-noise ratio of $\approx 400$). We use this reference star to test and calibrate three different PSF models. 

We first consider the PSF model produced by the Tiny Tim algorithm \citep{Krist95}, which uses a model of the HST optics and camera response to derive a model PSF. The HST focus model\footnote{http://www.stsci.edu/hst/observatory/focus/FocusModel} indicates the HST WFC3 observations of NGC~3156 were obtained in sub-optimal conditions; the secondary mirror despace is predicted to be $-8$~micron. Indeed, when this focus offset is used instead of the default value (i.e, no offset), the reduced $\chi_{r}^{2}$ of the Tiny Tim model for the reference star decreases from 34 to 21. Next we consider an empirical estimate of the typical WFC3 PSF based on an observation of the core of Omega Centauri \citep{2015wfc..rept....8A}. This ``Library'' PSF model is available\footnote{http://www.stsci.edu/hst/wfc3/analysis/PSF} for 28 different locations on each of the two WFC3 chips and for a large number of filters (for the F475W observations we use the Library PSF of the F438W filter). For WFC3 images, an empirically derived PSF model generally performs better than the Tiny Tim model \citep[see e.g.,][]{2012ApJS..203...24V}. To account for the focus difference between our images and the image used to derive the PSF library, we convolved the Library PSF at the location of the reference star with a Gaussian kernel that minimizes the residuals between this model and the star. The FWHM of the Library PSF is increased from 1.7 to 2.2 pixel by this step and the final reduced $\chi^{2}$ is 17. Finally, we constructed a PSF from the reference star directly by fitting a superposition of nine Gaussian profiles to the image of this star (the number of Gaussians used in this fit is simply set by by the requirement that the $\chi^{2}$ of this PSF model remains unchanged with the addition of the next Gaussian profile, which happens at $\chi_{r}^{2}=6.8$). 

While the PSF library allows for a correction of the spatial variations of the PSF, for our observations this variation is likely to be smaller than the change of the PSF due to the difference in focus between our observations and the mean focus of the Library PSF. We therefore anticipate that the Gaussian model derived for the reference star provides the best estimate of the true PSF at the location of the galaxy. However, as shown in the next section, all three PSF models that we considered yield very similar surface brightness profile parameters.

\subsubsection{Parameter inference}
We use \verb galfit  \citep{Peng+02} 
to fit a parametrized surface brightness profile to the image. This profile is convolved with each of the PSF models described above. We focus particularly on the inner 100~pc, as regions outside this have no influence on the TDE rate. Motivated by the results of ATLAS3D \citep{Krajno+13}, who showed that a power law provides a good description of the surface brightness profile on $\sim 10^{2}$~pc scales, we fit a broken power law to this inner region. This broken power law is equivalent to a Nuker profile (Eq.~\ref{eq:nuker}) with a large smoothing value ($\alpha\gg1$).

We use the F475W and F555W filters, which provide the highest-resolution view of the galaxy. The F475W and F555W observations were each obtained in a single orbit using two individual exposures (``dithers''), and the integration times were $2\times 370$~s and $2 \times 348$~s, respectively. 
Instead of co-adding (``drizzling'') the two frames, we use the individual flat-fielded frames (the \verb _flt  images) because the Tiny Tim and Library PSF model cannot be used for the ``drizzled'' image products.  
We estimate the parameters of the surface brightness profile by fitting a single profile to both frames simultaneously (i.e., we paste both frames to each other and treat this as one image in the \verb galfit  analysis). The results are summarized in Table~\ref{tab:matrix}. 

For each of the two filters and three PSF models we consider, the best-fit parameters reproduce the steep slope reported by  \citep{Krajno+13}, but flatten it to a ``core'' of inner slope $\gamma \approx 1.2$ at a radius of $R_{\rm b} \approx 5~{\rm pixels}$ or about $ 20~{\rm pc}$. This new profile, and its residuals, are shown in Fig. 2(b).  For all the different PSF models and filters we considered (Table~\ref{tab:matrix}) the inner slope is between 1.15 and 1.31. For a given PSF model, a change of the inner slope of only $\pm 0.05$ (with respect to the best-fit value) increases the $\chi^{2}_{r}$ by at least 2 (see Table~\ref{tab:chi2}). Hence the statistical uncertainty on the inner slope is smaller than the systematic uncertainty. 

\begin{table}[b]
\caption{Variation of the Inner Slope}\label{tab:chi2}
\begin{tabular}{ll |  l l}
Filter     &  PSF model & $\{I_{b},~R_{b},~\beta,~\gamma\}$    & $\chi^{2}/{\rm dof}$  \\
\hline\hline
F555W &   Library & $\{14.60,~4.33,~1.66,~1.17\}$ 					& $66.3$ \\
F555W &   Library & $\mathbf{\{14.94,~5.69,~1.69,~1.22\}}$  & $\mathbf{63.93}$ \\
F555W &   Library & $\{15.26,~6.38,~1.71,~1.27\}$ 					& $68.7$ \\
\hline
F555W &   Single star  & $\{14.23,~3.48,~1.68,~1.05\}$ 							& $64.0$ \\
F555W &   Single star & $\mathbf{\{14.71,~4.64,~1.71,~1.15\}}$ 		& $\mathbf{62.3}$\\
F555W &   Single star & $\{14.96,~5.39,~1.73,~1.20\}$ 							& $65.9$ \\

\end{tabular}	
\tablecomments{The best-fit parameters of the surface brightness profile (boldface), plus the parameters obtained for two sets when keeping the inner slope ($\gamma$) fixed at $\pm 0.05$ of the best-fit value. The surface brightness at the break radius $I_{\rm b}$ is given in AB mag$~{\rm arcsec}^{-2}$; the break radius $R_{\rm b}$ is given in pixels.  This small change to the inner slope leads to a large $\Delta \chi^{2}$, implying that the statistical uncertainty on the inner slope is smaller than the systematic uncertainty. }
\end{table}

One possible caveat to the above analysis concerns AGN activity in NGC~3156, as we have neglected any unresolved emission from the central SMBH.  The SDSS spectrum of NGC~3156 exhibits strong forbidden emission lines, and the absence of broad lines suggests that it is a type~II Seyfert galaxy. A lack of broad line emission implies that our line of sight to the central accretion disk is obscured; this is corroborated by the X-ray upper limit from ROSAT \citep{Voges99} observations ($L_{0.1-2.4~{kev}}<5 \times10^{39}\,{\rm erg}\,{\rm s}^{-1}$, 90\% confidence upper limit), which is a factor of $\sim 10$ lower than the expected X-ray flux for an unobscured source \cite[e.g.,][]{Hopkins07}. At least 95\% of type~II AGNs have gaseous absorbing column densities in excess of $N_{H}>10^{22}\,{\rm cm}^{-2}$ \citep{1999ApJ...522..157R}, which translates to a visual extinction of at least 5~mag. Hence, the contribution of AGN emission to the {\it HST} surface brightness profile is expected to be negligible. 

The F225W$-$F555W color in an aperture of 2 pixels is 3.2 and is similar to the color seen throughout the galaxy (aside from variations in visible dust lanes). The NUV$-$r color places this galaxy -- and its nucleus -- in the ``green valley'' \citep{2007ApJS..173..293W}, and a lack of strong color gradients toward the inner few pixels further shows that our surface brightness profiles suffer little contamination from unresolved AGN light. To estimate the maximum AGN contribution we consider the (unlikely) scenario in which all of the observed unresolved NUV emission is due to an AGN and we convert this NUV flux to the F475W band using the mean optical spectrum of unobscured AGNs, $F_{\nu}\propto \nu^{-0.4}$ \citep{VandenBerk2001}. When we include this estimate of the AGN light into our model for the surface brightness profile, the inferred inner slope ($\gamma$) is smaller by 0.04 compared to the power-law index measured without this contribution. This conservative estimate of the influence of AGN emission to the inner slope is similar to the systematic uncertainty due to the PSF model. 

\section{NGC~3156: Tidal Disruption Event Rate}
\label{sec:rate}

To quantify the TDE rate in NGC~3156, we employ the formalism of \citet{WanMer04}.  Specifically, we deproject $I(R)$ into a 3D density profile $\rho(r)$, assuming spherical symmetry.  To accommodate theoretically motivated changes to the profile  below the {\it HST} resolution limit, we introduce a softened surface brightness profile analogous to the classical Nuker law:
\begin{align}
I(R) =& 2^{(\beta-\gamma)/\alpha}I_{\rm b}'\left(\frac{R_{\rm c}}{R} \right)^\delta \left[1+\left(\frac{R}{R_{\rm c}} \right)^{\alpha_{\rm c}} \right]^{(\delta-\gamma)/\alpha_{\rm c}} \\
&\times \left[1+\left(\frac{R}{R_{\rm b}} \right)^\alpha \right]^{(\gamma-\beta)/\alpha}, \notag
\end{align}
In this ``double Nuker'' profile, $I_{\rm b}' = I_{\rm b}(R_{\rm b}/R_{\rm c})^\delta(1+(R_{\rm b}/R_{\rm c})^{\alpha_{\rm c}})^{(\gamma-\delta)/\alpha_{\rm c}}$.  We fix $\alpha=\alpha_{\rm c}=10$, which produces an $I(R)$ profile very similar to an infinitely sharp break ($\alpha_{\rm c}=\infty$), but that avoids unwanted deprojection errors.  The outer power-law index $\beta$, the intermediate power-law index $\gamma$, and the outer break radius $R_{\rm b}$ are all fitted to the observed light from the innermost $100~{\rm pc}$ in NGC 3156; the inner power-law index $\delta$ and break radius $R_{\rm c}$ are sub-resolution parameters that are set by the theoretical considerations we detail below.

We use $\rho(r)$ to calculate the gravitational potential $\psi(r)$ and then the stellar distribution function $f(\varepsilon)$, assuming isotropic velocities\footnote{Detailed dynamical modeling of NGC 3156 indicates a global bias toward modestly radial orbits \citep{Cappel+07}; if such a bias holds down to very small radii, this would increase the true TDE rate above our isotropic calculation.  A detailed anisotropic modeling of NGC 3156 is beyond the scope of this paper, and in any case our goal is to benchmark this rate calculation against TDE rates in large samples of galaxies, which have always been computed under the assumption of velocity isotropy \citep{WanMer04, StoMet14}.}.  Here both the potential $\psi$ and specific orbital energy $\varepsilon$ are written in stellar dynamics notation (bound orbits are positive).  The distribution function is used to compute orbit-averaged diffusion coefficients $\bar{\mu}(\varepsilon)$ due to two-body relaxation, which in turn provide the flux of stars into the loss cone per energy bin per time, $\mathcal{F}(\varepsilon)$.  Finally, we multiply $\mathcal{F}(\varepsilon)$ by a correction factor accounting for a mass spectrum of stars\footnote{In this paper, we assume a Kroupa initial mass function \citep{Kroupa01} truncated at a maximum mass of $m_\star = 3M_\odot$.  Compared to a single-mass distribution, TDE rates are increased by the larger number of stars present in the mass function, but decreased by the reduction in diffusion coefficients; the net effect is a modest increase, by a factor of $2.22$, over the equivalent calculation where all stars possess $m_\star = M_\odot$.  We neglect the changing tidal radius for stars of different masses because it varies little across the lower main sequence and alters the loss cone flux $\mathcal{F}(\varepsilon)$ by a factor that is at most logarithmic in the ratio of tidal radii.}, as described in Appendix A of \citet{MagTre99}, and compute the total rate $\dot{N}=\int \mathcal{F}(\varepsilon){\rm d}\varepsilon$.  We refer the reader to the original literature for a detailed summary of this procedure \citep{MagTre99, WanMer04}.

The primary inputs to our calculation are the surface brightness profile $I(R)$; the SMBH mass $M_\bullet$; and the mass-to-light ratio $\Upsilon$.  Using the HST photometry (i.e., the F225W, F475W, F555W, and F814W filters), we find that for the F475W filter, the galaxy-averaged value of $\Upsilon=1.58$ shrinks to $\Upsilon=0.485$ for the central 50 pc; this is consistent with the radial color gradients found in a broader investigation of E+A galaxies \citep{Pracy+12, Pracy+13}, and indicates that star formation in NGC~3156 was preferentially concentrated in its nucleus.  For the F555W filter, we find $\Upsilon=0.496$ in the central 50 pc.  

To test the sensitivity of our calculated TDE rate to uncertainties in the observations, we consider six surface brightness profiles $I(R)$.  In scenarios A1, A2, and A3, we use the F555W filter and a PSF modeled with Tiny Tim, the calibration star, and the empirical Library PSF model, respectively.  In scenarios B1, B2, and B3, we use the same respective PSF models for the F475W filter.  In these six fiducial scenarios, we find intermediate power-law slopes $1.15 \le \gamma \le 1.31$.  These slopes are extreme outliers compared with the inner slopes of most other observed galaxies \citep{Lauer+07a}, and in many cases will produce a formally divergent TDE rate: if $\delta=5/4$, equal logarithmic intervals in energy space will contribute equally to the TDE rate inside the SMBH influence radius, and if $\delta\ge 5/4$, the TDE rate diverges when integrated to $\varepsilon=\infty$ \citep{SyeUlm99}.

We therefore fix $\delta=3/4$, the value typical for a relaxed, idealized stellar system in the SMBH influence radius \citep{BahWol76}.  We set the location of the transition to be $R_{\rm c}=r_{\rm BW}$, the ``Bahcall-Wolf'' radius where the relaxation time is equal to  $t_{\rm age}$, the age of the system:
\begin{equation}
r_{\rm BW}^{3/2-\Gamma} = \frac{0.34M_\bullet^{3/2}\langle m_\star \rangle}{G^{1/2}\langle m_\star^2 \rangle t_{\rm age} \rho(r_0)r_0^\Gamma \ln \Lambda}.
\end{equation}
Here $\Gamma = \gamma+1$ is the power-law slope of the inner 3D density profile\footnote{We note that inside the SMBH sphere of influence, systems with $\Gamma > 3/2$ relax from the inside out, while those with $\Gamma < 3/2$ relax from the outside in.  This formula fails to apply if it predicts $r_{\rm BW} > r_{\rm infl}$, but this does not occur for our fiducial parameter choices and $t_{\rm age} \le 10^9~{\rm yr}$.}.  In this equation $\langle m_\star \rangle$ and $\langle m_\star^2 \rangle$ are the first and second moments of the stellar present day mass function (PDMF).  We take a Kroupa initial mass function \citep{Kroupa01} and truncate it at a maximum $m_\star = 3M_\odot$ to approximate the PDMF of a post-starburst galaxy with a large population of A stars.  The reference radius $r_0$ is any radius that satisfies $r_{\rm BW} < r_0 < R_{\rm b}$, and we take the Coulomb logarithm to be $\Lambda \equiv 0.4 M_\bullet / \langle m_\star \rangle$.  We conservatively take $t_{\rm age} = 10^9~{\rm yr}$, which increases $r_{\rm BW}$ and decreases the TDE rate relative to younger nuclear starbursts.

\begin{figure}
\includegraphics[width=85mm]{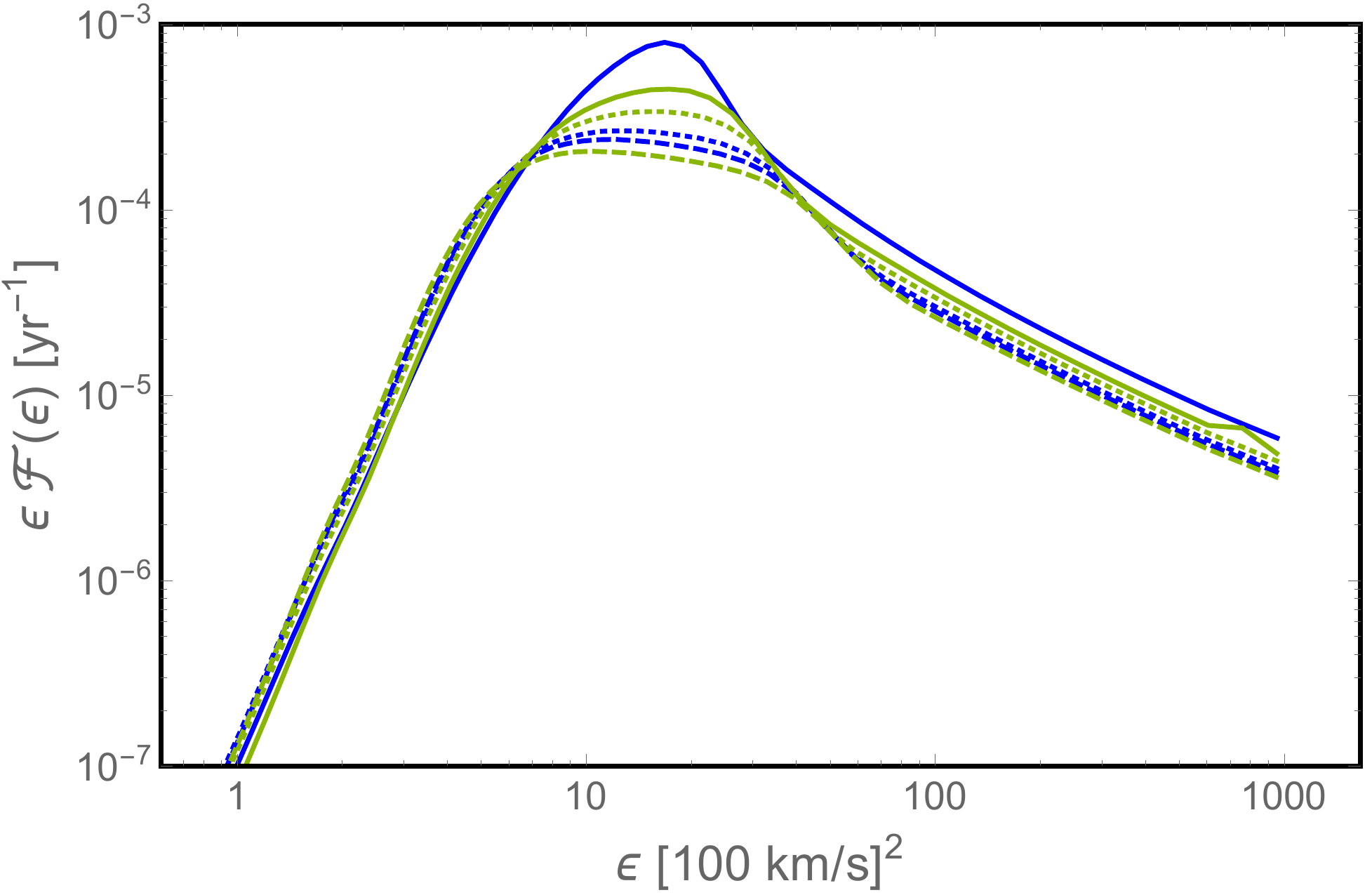}
\caption{Flux of stars into the SMBH loss cone in NGC~3156.  The loss cone flux $\mathcal{F}(\epsilon)$ is a function of specific orbital energy $\epsilon$, and has units of per time per specific energy.  The blue and green lines show the F555W and F475W filters, respectively, while the dotted, dashed, and solid lines correspond to the Tiny Tim, calibration star, and Anderson PSF models.  The flux curves depend only weakly on the choice of filter or PSF model.}
\label{fig:LCFlux}
\end{figure}

The results for all scenarios are shown in Fig.~\ref{fig:LCFlux}.  Generally speaking, we find that the TDE rate in NGC~3156 is quite insensitive to the choice of PSF model or filter.  The TDE rates in scenarios A1, A2, A3, B1, B2, and B3 are, respectively, $1.4\times10^{-3}~{\rm yr}^{-1}$, $1.1\times10^{-3}~{\rm yr}^{-1}$, $1.7\times10^{-3}~{\rm yr}^{-1}$, $1.3\times10^{-3}~{\rm yr}^{-1}$, $1.2\times10^{-3}~{\rm yr}^{-1}$, and $2.2\times10^{-3}~{\rm yr}^{-1}$.

\begin{figure}
\includegraphics[width=85mm]{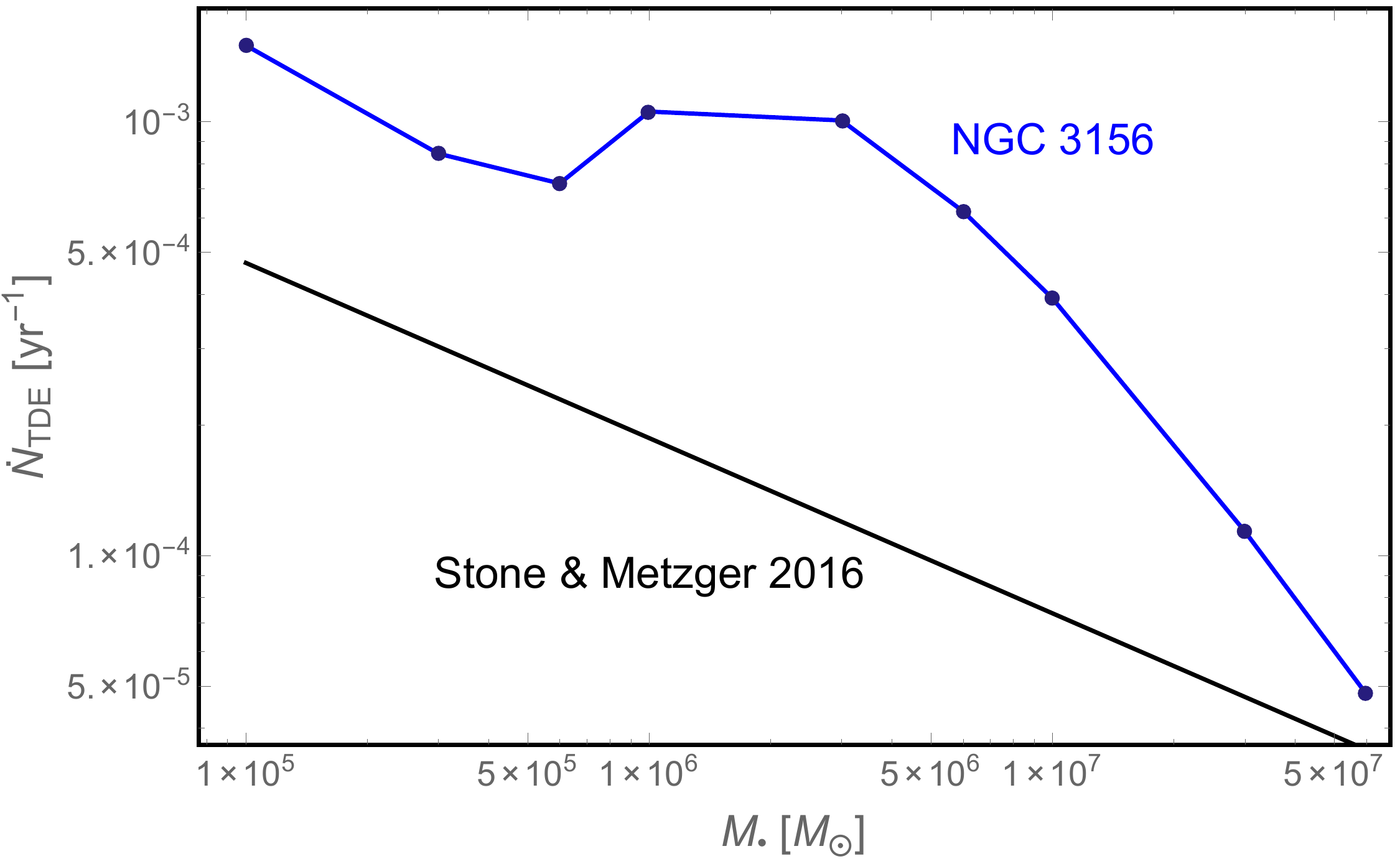}
\caption{Tidal disruption rates $\dot{N}$ as a function of SMBH mass $M_\bullet$.  The blue curve shows model A3 for NGC 3156 and is fairly representative of all six of our models for $I(R)$.  The black curve shows the power law best-fit for a large galaxy sample obtained from \citet{StoMet14}.  }
\label{fig:rateVMass}
\end{figure}

Our results are more sensitive to the choice of $M_\bullet$, but not within the range estimated from galaxy scaling relations ($1\times 10^6 \lesssim M_\bullet/M_\odot \lesssim 3\times 10^6$).  If the true value of $M_\bullet$ falls significantly above or below these values, the TDE rate will decrease from its fiducial $\dot{N}\sim 1\times 10^{-3}~{\rm yr}^{-1}$ value.  At higher masses, this occurs because of the changing position of the phase space critical radius; at lower masses, this occurs because $r_{\rm BW}$ is growing larger.  We illustrate $\dot{N}(M_\bullet)$ in Fig.~\ref{fig:rateVMass}, and show that for fiducial SMBH masses, it is an order of magnitude higher than that of typical galaxies.

By far the largest uncertainty in this calculation, however, is the inward extrapolation of the $I(R) \propto R^{-\gamma}$ power law, which fits  scales from $R_{\rm b} \approx 20~{\rm pc}$ down to the resolution limit at $R \approx 4~{\rm pc}$.  In \citet{StoMet14}, most SMBHs in this mass range have critical radii that are unresolved by factors of a few; the greater steepness of the surface brightness profile in NGC 3156 means that its critical radius is underresolved by over an order of magnitude (assuming an extrapolation down to the Bahcall-Wolf radius, which varies between $0.02~{\rm pc}$ and $0.09~{\rm pc}$ in the six fiducial models we consider).  In performing this extrapolation we follow the procedure adopted in other TDE rate calculations for low-mass galaxies \citep{MagTre99, WanMer04, StoMet14}, but we emphasize that if the density profile in NGC 3156 turns over at a radius $R_{\rm c}$ such that $r_{\rm BW} \ll R_{\rm c} < R_{\rm res}$, it is possible to bring the TDE rate down to $\dot{N} \approx 2 \times 10^{-4}~{\rm yr}^{-1}$, a more typical value for an SMBH of this size.  For example, in model A3, $r_{\rm BW} = 0.05~{\rm pc}$, but if we manually force $R_{\rm c}$ to larger values, $\dot{N}$ falls below $1\times 10^{-3}~{\rm yr}^{-1}$ for $R_{\rm c} \gtrsim 0.2~{\rm pc}$ and becomes as low as $4\times 10^{-4}~{\rm yr}^{-1}$ for $R_{\rm c} = 0.8~{\rm pc}$.  Larger values of $R_{\rm c}$ cause deficits of order unity in the unresolved light and are therefore ruled out.

\section{Discussion}
\label{sec:discussion}

NGC~3156 possesses an extraordinarily steep density profile on scales of $\approx 50~{\rm pc}$.  A naive extrapolation of NGC~3156's inner surface brightness profile from this region would give an enormously overdense stellar population (and a correspondingly large TDE rate), but we have shown that a turnover on scales $R_{\rm b} \approx 20~{\rm pc}$ flattens $I(R)$ to a power-law index of $\gamma \approx 1.15-1.3$.  These values are still larger than the power-law index seen in any of the 144 galaxies considered in \citet{StoMet14}, and suggest that NGC~3156 may be a TDE factory.

Our numerical rate calculations bear out this suggestion: the fiducial TDE rate of $\dot{N} \approx 1\times 10^{-3}~{\rm yr}^{-1}$ is in agreement with the observationally inferred TDE rate in E+A galaxies \citep{French+16}.  When one calculates TDE rates in typical early-type galaxies of comparable SMBH mass through the procedure employed in this paper, the resulting $\dot{N} \sim 10^{-4}~{\rm yr}^{-1}$, an order of magnitude lower than that in NGC 3156.  While the analysis of \citet{French+16} suggests that E+A galaxies should have TDE rates that are two orders of magnitude greater than those in typical galaxies, we note that the extra order of magnitude is likely tied to the {\it underproduction} of TDEs in standard early-type galaxies \citep{StoMet14}.  Overall, our fiducial models for NGC~3156 appear entirely consistent with the hypothesis that E+A galaxies overproduce TDEs because central overdensities lead to short two-body relaxation timescales.  The one major caveat in our work is that we have had to extrapolate the observed $I(R)$ profile well below the {\it HST} resolution limit.  This is the standard procedure in other theoretical TDE rate calculations that are the primary point of comparison for our results on NGC 3156, but is nonetheless a limitation of our modeling.

Unfortunately, current observations do not constrain alternative hypotheses seeking to enhance TDE rates in post-starburst galaxies.  There is no clearly flattened core indicative of a post-starburst SMBH binary inspiral, although we cannot rule out flattening inside the {\it HST} resolution limit.  Theory predicts that the inspiral of an SMBH binary would excavate a mass deficit comparable to $M_\bullet$ \citep{Merrit07}, but if we crudely estimate a mass deficit by differencing the enclosed mass at 20 pc for both scenario A2 and an unbroken power law of slope $\beta=1.71$, we find $M_{\rm def} = 8.2\times 10^7M_\odot \gg M_\bullet$.  This number is far too large to constrain a recent SMBH binary inspiral, and likely indicates not scouring, but a decreasing star formation efficiency inward of $R_{\rm b}$.

An alternative explanation for the E+A enhancement could arise from a prevalence of highly aspherical stellar potentials in the nuclei of this galaxy type.  Because stellar orbits do not fully conserve angular momentum in such potentials, their presence enhances TDE rates significantly \citep{MagTre99}.  The inner isophotes of NGC 3156 are indeed non-circular, with an average ellipticity of $\epsilon = 0.4$ in the inner $17$ pc.  Such a value indicates some axisymmetry, but not an unusually high amount \citep{Lauer+05}, and in any case axisymmetry alone will only enhance TDE rates by factors of a few \citep{VasMer13}.  Triaxial geometries can produce much larger enhancements \citep{MerPoo04} and are therefore a more plausible explanation for the E+A enhancement.  However, without detailed kinematic data, we are unable to estimate the nuclear triaxiality of NGC 3156. 

We also note two final points of interest that are not directly related to our main investigation.
\begin{itemize}
\item Our best-fit break radius, $R_{\rm b}\approx 20~{\rm pc}$, is comparable to the tidal radius for a dense giant molecular cloud (GMC).  If we take a GMC density of $10^4~{\rm cm}^{-3}$ and our scenario A2, the mean density of the enclosed mass (stars and SMBH) equals that of the cloud at $\approx 30~{\rm pc}$.  Such GMC densities are typical for the central $200$ pc of the Milky Way \citep{MorSer96}, and the stellar density turnover we infer provides tentative evidence that tidal shearing effectively chokes star formation in the starbursts that produce E+A galaxies.  Perhaps a different mode of star formation, with a different efficiency, produced the young stars interior to this radius \citep{Thomps+05}.  This flattening is not a Bahcall-Wolf cusp reflecting stellar dynamical equilibrium; the relaxation times at this radius are $\gtrsim 10^{11}~{\rm yr}$ in all of our models.
\item The TDE rate we estimated following careful PSF deconvolution differed by multiple orders of magnitude from that which would have been calculated from a global Nuker law fit.  This suggests that caution should be used in interpreting TDE rate calculations \citep{MagTre99, WanMer04, StoMet14} that use globally fitted $I(R)$ parametrizations.  More customized surface brightness parametrizations (or, perhaps, a nonparametric calculation) may be necessary to more accurately capture stellar light profiles near the {\it HST} resolution limit.  If most galaxies possess central $I(R)$ turnovers similar to that of NGC 3156, this could address the ``rate discrepancy'' identified in \citet{StoMet14}.  However, we note that the tension between observationally inferred (low) and dynamically predicted (high) TDE rates could also be worsened by this type of detailed modeling, as many of the Nuker fits used in \citet{StoMet14} specifically excluded light overdensities from nuclear star clusters (T. Lauer, private communication).  The addition of these dense star clusters via nonparametric modeling would increase TDE rates in their host galaxies.
\end{itemize}

We have shown that NGC 3156 possesses an unusually steep surface brightness profile down to the {\it HST} resolution limit.  If this surface brightness profile is extrapolated inward, we find a TDE rate of $\dot{N}\approx 1\times 10^{-3}~{\rm yr}^{-1}$, consistent with observationally inferred TDE rates in E+A galaxies.  This number is an order of magnitude greater than the typical rates calculated in other low-mass galaxies using analogous extrapolations.  Future {\it HST} photometry of other nearby E+As would allow this exercise to be repeated with a larger sample size, statistically testing the overdensity hypothesis.  Because galaxies with steeper density cusps (and higher TDE rates) have loss cone flux curves peaking at smaller radii, the overdensity hypothesis will be easier to falsify than to validate.  Until then, the unusually steep surface brightness profile of NGC 3156 provides preliminary evidence that the unusual host galaxy preference of TDEs is tied to nuclear stellar overdensities created in the starbursts that produce E+A and Balmer-strong galaxies.

\begin{table*}
\centering
\caption{Matrix of inferred parameters}
\begin{tabular}{ll||lll}
Filter						&Dither 			&Tiny Tim 				 & Single star										&Library\\
\hline\hline
F475W&1			&$\{15.07,~5.71,~1.97,~1.15\}$&$\{14.80,~4.77,~1.85,~1.12\}$&$\{15.47,~6.79,~1.79,~1.29\}$\\
F475W&2			&$\{15.22,~6.15,~1.97,~1.21\}$&$\{15.00,~5.13,~1.84,~1.19\}$&$\{15.68,~7.61,~1.78,~1.33\}$\\
F475W&both		&$\{14.98,~5.36,~1.90,~1.16\}$&$\{14.78,~4.61,~1.80,~1.15\}$&$\{15.39,~6.40,~1.70,~1.31\}$\\
\hline
F555W&1			&$\{15.10,~6.15,~1.86,~1.22\}$&$\{14.84,~5.05,~1.73,~1.18\}$&$\{15.11,~5.90,~1.70,~1.25\}$\\
F555W&2			&$\{15.05,~6.01,~1.88,~1.19\}$&$\{14.82,~5.03,~1.76,~1.15\}$&$\{15.07,~5.78,~1.74,~1.21\}$\\
F555W&both 	&$\{14.85,~5.32,~1.81,~1.19\}$&$\{14.71,~4.64,~1.71,~1.15\}$&$\{14.94,~5.69,~1.69,~1.22\}$\\
\end{tabular}
\tablecomments{Parameters of our broken power-law model for the surface brightness in the inner 100~pc of NGC~3156. Each set of elements lists $\{I_{b},~R_{b},~\beta,~\gamma\}$ (see Eq.~\ref{eq:nuker}), with the same units as in Table 1. We show the results obtained for each of the three different PSF models, two different filters (F475W and F555W), and the two images (``dithers'') for each filter.}\label{tab:matrix}
\end{table*}

\section*{Acknowledgments}
We thank Jay Anderson, Decker French, Andrew Fruchter, Davor Krajnovi\'{c}, Julian Krolik, Brian Metzger, Elena Sabbi, Gregory Snyder, and Ann Zabludoff for useful discussions.  This work is based on observations made with the NASA/ESA Hubble Space Telescope, and obtained from the Hubble Legacy Archive, which is a collaboration between the Space Telescope Science Institute (STScI/NASA), the Space Telescope European Coordinating Facility (ST-ECF/ESA) and the Canadian Astronomy Data Centre (CADC/NRC/CSA).  Financial support was provided to N.C.S. by NASA through Einstein Postdoctoral Fellowship Award Number PF5-160145, and to S.V.V. through Hubble Postdoctoral Fellowship Award Number HST-HF2-51350.

\bibliographystyle{apj}
\bibliography{ms}

\begin{thebibliography}{}
\expandafter\ifx\csname natexlab\endcsname\relax\def\natexlab#1{#1}\fi

\bibitem[{REV(????)}]{REVTEX41Control}
 ????

\bibitem[{08(1)}]{apsrev41Control}
08. 1

\bibitem[{{Adelman-McCarthy} {et~al.}(2008){Adelman-McCarthy}, {Ag{\"u}eros},
  {Allam}, {Allende Prieto}, {Anderson}, {Anderson}, {Annis}, {Bahcall},
  {Bailer-Jones}, {Baldry}, {Barentine}, {Bassett}, {Becker}, {Beers}, {Bell},
  {Berlind}, {Bernardi}, {Blanton}, {Bochanski}, {Boroski}, {Brinchmann},
  {Brinkmann}, {Brunner}, {Budav{\'a}ri}, {Carliles}, {Carr}, {Castander},
  {Cinabro}, {Cool}, {Covey}, {Csabai}, {Cunha}, {Davenport}, {Dilday}, {Doi},
  {Eisenstein}, {Evans}, {Fan}, {Finkbeiner}, {Friedman}, {Frieman},
  {Fukugita}, {G{\"a}nsicke}, {Gates}, {Gillespie}, {Glazebrook}, {Gray},
  {Grebel}, {Gunn}, {Gurbani}, {Hall}, {Harding}, {Harvanek}, {Hawley},
  {Hayes}, {Heckman}, {Hendry}, {Hindsley}, {Hirata}, {Hogan}, {Hogg}, {Hyde},
  {Ichikawa}, {Ivezi{\'c}}, {Jester}, {Johnson}, {Jorgensen}, {Juri{\'c}},
  {Kent}, {Kessler}, {Kleinman}, {Knapp}, {Kron}, {Krzesinski}, {Kuropatkin},
  {Lamb}, {Lampeitl}, {Lebedeva}, {Lee}, {Leger}, {L{\'e}pine}, {Lima}, {Lin},
  {Long}, {Loomis}, {Loveday}, {Lupton}, {Malanushenko}, {Malanushenko},
  {Mandelbaum}, {Margon}, {Marriner}, {Mart{\'{\i}}nez-Delgado}, {Matsubara},
  {McGehee}, {McKay}, {Meiksin}, {Morrison}, {Munn}, {Nakajima}, {Neilsen},
  {Newberg}, {Nichol}, {Nicinski}, {Nieto-Santisteban}, {Nitta}, {Okamura},
  {Owen}, {Oyaizu}, {Padmanabhan}, {Pan}, {Park}, {Peoples}, {Pier}, {Pope},
  {Purger}, {Raddick}, {Re Fiorentin}, {Richards}, {Richmond}, {Riess}, {Rix},
  {Rockosi}, {Sako}, {Schlegel}, {Schneider}, {Schreiber}, {Schwope}, {Seljak},
  {Sesar}, {Sheldon}, {Shimasaku}, {Sivarani}, {Smith}, {Snedden}, {Steinmetz},
  {Strauss}, {SubbaRao}, {Suto}, {Szalay}, {Szapudi}, {Szkody}, {Tegmark},
  {Thakar}, {Tremonti}, {Tucker}, {Uomoto}, {Vanden Berk}, {Vandenberg},
  {Vidrih}, {Vogeley}, {Voges}, {Vogt}, {Wadadekar}, {Weinberg}, {West},
  {White}, {Wilhite}, {Yanny}, {Yocum}, {York}, {Zehavi}, \&
  {Zucker}}]{Adelma+08}
{Adelman-McCarthy}, J.~K., {Ag{\"u}eros}, M.~A., {Allam}, S.~S., {et~al.} 2008,
  \apjs, 175, 297

\bibitem[{{Anderson} {et~al.}(2015){Anderson}, {Bourque}, {Sahu}, {Sabbit}, \&
  {Viana}}]{2015wfc..rept....8A}
{Anderson}, J., {Bourque}, M., {Sahu}, K., {Sabbit}, E., \& {Viana}, A. 2015,
  {A Study of the Time Variability of the PSF in F606W Images taken with the
  WFC3/UVIS}, Tech. rep.

\bibitem[{{Arcavi} {et~al.}(2014)}]{Arcavi+14}
{Arcavi}, I., {et~al.} 2014, \apj, 793, 38

\bibitem[{{Bade} {et~al.}(1996){Bade}, {Komossa}, \& {Dahlem}}]{Bade+96}
{Bade}, N., {Komossa}, S., \& {Dahlem}, M. 1996, \aap, 309, L35

\bibitem[{{Bahcall} \& {Wolf}(1976)}]{BahWol76}
{Bahcall}, J.~N., \& {Wolf}, R.~A. 1976, \apj, 209, 214

\bibitem[{{Blanton} \& {Roweis}(2007)}]{BlaRow07}
{Blanton}, M.~R., \& {Roweis}, S. 2007, \aj, 133, 734

\bibitem[{{Blanton} {et~al.}(2005){Blanton}, {Schlegel}, {Strauss},
  {Brinkmann}, {Finkbeiner}, {Fukugita}, {Gunn}, {Hogg}, {Ivezi{\'c}}, {Knapp},
  {Lupton}, {Munn}, {Schneider}, {Tegmark}, \& {Zehavi}}]{Blanto+05}
{Blanton}, M.~R., {Schlegel}, D.~J., {Strauss}, M.~A., {et~al.} 2005, \aj, 129,
  2562

\bibitem[{{Cappellari} {et~al.}(2006){Cappellari}, {Bacon}, {Bureau}, {Damen},
  {Davies}, {de Zeeuw}, {Emsellem}, {Falc{\'o}n-Barroso}, {Krajnovi{\'c}},
  {Kuntschner}, {McDermid}, {Peletier}, {Sarzi}, {van den Bosch}, \& {van de
  Ven}}]{Cappel+06}
{Cappellari}, M., {Bacon}, R., {Bureau}, M., {et~al.} 2006, \mnras, 366, 1126

\bibitem[{{Cappellari} {et~al.}(2007){Cappellari}, {Emsellem}, {Bacon},
  {Bureau}, {Davies}, {de Zeeuw}, {Falc{\'o}n-Barroso}, {Krajnovi{\'c}},
  {Kuntschner}, {McDermid}, {Peletier}, {Sarzi}, {van den Bosch}, \& {van de
  Ven}}]{Cappel+07}
{Cappellari}, M., {Emsellem}, E., {Bacon}, R., {et~al.} 2007, \mnras, 379, 418

\bibitem[{{Cappellari} {et~al.}(2013){Cappellari}, {Scott}, {Alatalo}, {Blitz},
  {Bois}, {Bournaud}, {Bureau}, {Crocker}, {Davies}, {Davis}, {de Zeeuw},
  {Duc}, {Emsellem}, {Khochfar}, {Krajnovi{\'c}}, {Kuntschner}, {McDermid},
  {Morganti}, {Naab}, {Oosterloo}, {Sarzi}, {Serra}, {Weijmans}, \&
  {Young}}]{Cappel+13}
{Cappellari}, M., {Scott}, N., {Alatalo}, K., {et~al.} 2013, \mnras, 432, 1709

\bibitem[{{Chen} {et~al.}(2011){Chen}, {Sesana}, {Madau}, \& {Liu}}]{Chen+11}
{Chen}, X., {Sesana}, A., {Madau}, P., \& {Liu}, F.~K. 2011, \apj, 729, 13

\bibitem[{{Chornock} {et~al.}(2014)}]{Chorno+14}
{Chornock}, R., {et~al.} 2014, \apj, 780, 44

\bibitem[{{de Vaucouleurs} {et~al.}(1991){de Vaucouleurs}, {de Vaucouleurs},
  {Corwin}, {Buta}, {Paturel}, \& {Fouqu{\'e}}}]{deVauco+91}
{de Vaucouleurs}, G., {de Vaucouleurs}, A., {Corwin}, Jr., H.~G., {et~al.}
  1991, {Third Reference Catalogue of Bright Galaxies. Volume I: Explanations
  and references. Volume II: Data for galaxies between 0$^{h}$ and 12$^{h}$.
  Volume III: Data for galaxies between 12$^{h}$ and 24$^{h}$.}

\bibitem[{{Donley} {et~al.}(2002){Donley}, {Brandt}, {Eracleous}, \&
  {Boller}}]{Donley+02}
{Donley}, J.~L., {Brandt}, W.~N., {Eracleous}, M., \& {Boller}, T. 2002, \aj,
  124, 1308

\bibitem[{{Esquej} {et~al.}(2007){Esquej}, {Saxton}, {Freyberg}, {Read},
  {Altieri}, {Sanchez-Portal}, \& {Hasinger}}]{Esquej+07}
{Esquej}, P., {Saxton}, R.~D., {Freyberg}, M.~J., {et~al.} 2007, \aap, 462, L49

\bibitem[{{Esquej} {et~al.}(2008){Esquej}, {Saxton}, {Komossa}, {Read},
  {Freyberg}, {Hasinger}, {Garc{\'{\i}}a-Hern{\'a}ndez}, {Lu}, {Rodriguez
  Zaur{\'{\i}}n}, {S{\'a}nchez-Portal}, \& {Zhou}}]{Esquej+08}
{Esquej}, P., {Saxton}, R.~D., {Komossa}, S., {et~al.} 2008, \aap, 489, 543

\bibitem[{{Frank} \& {Rees}(1976)}]{FraRee76}
{Frank}, J., \& {Rees}, M.~J. 1976, \mnras, 176, 633

\bibitem[{{French} {et~al.}(2016){French}, {Arcavi}, \&
  {Zabludoff}}]{French+16}
{French}, K.~D., {Arcavi}, I., \& {Zabludoff}, A. 2016, \apjl, 818, L21

\bibitem[{{Fukugita} {et~al.}(1996){Fukugita}, {Ichikawa}, {Gunn}, {Doi},
  {Shimasaku}, \& {Schneider}}]{Fukugi+96}
{Fukugita}, M., {Ichikawa}, T., {Gunn}, J.~E., {et~al.} 1996, \aj, 111, 1748

\bibitem[{{Gezari} {et~al.}(2006){Gezari}, {Martin}, {Milliard}, {Basa},
  {Halpern}, {Forster}, {Friedman}, {Morrissey}, {Neff}, {Schiminovich},
  {Seibert}, {Small}, \& {Wyder}}]{Gezari+06}
{Gezari}, S., {Martin}, D.~C., {Milliard}, B., {et~al.} 2006, \apjl, 653, L25

\bibitem[{{Gezari} {et~al.}(2008){Gezari}, {Basa}, {Martin}, {Bazin},
  {Forster}, {Milliard}, {Halpern}, {Friedman}, {Morrissey}, {Neff},
  {Schiminovich}, {Seibert}, {Small}, \& {Wyder}}]{Gezari+08}
{Gezari}, S., {Basa}, S., {Martin}, D.~C., {et~al.} 2008, \apj, 676, 944

\bibitem[{{Gezari} {et~al.}(2012)}]{Gezari+12}
{Gezari}, S., {et~al.} 2012, \nat, 485, 217

\bibitem[{{Hills}(1975)}]{Hills75}
{Hills}, J.~G. 1975, \nat, 254, 295

\bibitem[{{Holoien} {et~al.}(2014){Holoien}, {Prieto}, {Bersier}, {Kochanek},
  {Stanek}, {Shappee}, {Grupe}, {Basu}, {Beacom}, {Brimacombe}, {Brown},
  {Davis}, {Jencson}, {Pojmanski}, \& {Szczygiel}}]{Holoie+14}
{Holoien}, T.~W.-S., {Prieto}, J.~L., {Bersier}, D., {et~al.} 2014, ArXiv
  e-prints, arXiv:1405.1417

\bibitem[{{Holoien} {et~al.}(2016){Holoien}, {Kochanek}, {Prieto}, {Stanek},
  {Dong}, {Shappee}, {Grupe}, {Brown}, {Basu}, {Beacom}, {Bersier},
  {Brimacombe}, {Danilet}, {Falco}, {Guo}, {Jose}, {Herczeg}, {Long},
  {Pojmanski}, {Simonian}, {Szczygie{\l}}, {Thompson}, {Thorstensen}, {Wagner},
  \& {Wo{\'z}niak}}]{Holoie+16}
{Holoien}, T.~W.-S., {Kochanek}, C.~S., {Prieto}, J.~L., {et~al.} 2016, \mnras,
  455, 2918

\bibitem[{{Hopkins} {et~al.}(2007){Hopkins}, {Richards}, \&
  {Hernquist}}]{Hopkins07}
{Hopkins}, P.~F., {Richards}, G.~T., \& {Hernquist}, L. 2007, \apj, 654, 731

\bibitem[{{Ivanov} {et~al.}(2005){Ivanov}, {Polnarev}, \& {Saha}}]{Ivanov+05}
{Ivanov}, P.~B., {Polnarev}, A.~G., \& {Saha}, P. 2005, \mnras, 358, 1361

\bibitem[{{Karas} \& {{\v S}ubr}(2007)}]{KarSub07}
{Karas}, V., \& {{\v S}ubr}, L. 2007, \aap, 470, 11

\bibitem[{{Khabibullin} \& {Sazonov}(2014)}]{KhaSaz14}
{Khabibullin}, I., \& {Sazonov}, S. 2014, \mnras, 444, 1041

\bibitem[{{Komossa} \& {Greiner}(1999)}]{KomGre99}
{Komossa}, S., \& {Greiner}, J. 1999, \aap, 349, L45

\bibitem[{{Kormendy} \& {Ho}(2013)}]{KorHo13}
{Kormendy}, J., \& {Ho}, L.~C. 2013, \araa, 51, 511

\bibitem[{{Krajnovi{\'c}} {et~al.}(2013){Krajnovi{\'c}}, {Karick}, {Davies},
  {Naab}, {Sarzi}, {Emsellem}, {Cappellari}, {Serra}, {de Zeeuw}, {Scott},
  {McDermid}, {Weijmans}, {Davis}, {Alatalo}, {Blitz}, {Bois}, {Bureau}, \&
  {Bournaud}}]{Krajno+13}
{Krajnovi{\'c}}, D., {Karick}, A.~M., {Davies}, R.~L., {et~al.} 2013, \mnras,
  433, 2812

\bibitem[{{Krist}(1995)}]{Krist95}
{Krist}, J. 1995, in Astronomical Society of the Pacific Conference Series,
  Vol.~77, Astronomical Data Analysis Software and Systems IV, ed. R.~A.
  {Shaw}, H.~E. {Payne}, \& J.~J.~E. {Hayes}, 349

\bibitem[{{Kroupa}(2001)}]{Kroupa01}
{Kroupa}, P. 2001, \mnras, 322, 231

\bibitem[{{Lauer} {et~al.}(2005){Lauer}, {Faber}, {Gebhardt}, {Richstone},
  {Tremaine}, {Ajhar}, {Aller}, {Bender}, {Dressler}, {Filippenko}, {Green},
  {Grillmair}, {Ho}, {Kormendy}, {Magorrian}, {Pinkney}, \&
  {Siopis}}]{Lauer+05}
{Lauer}, T.~R., {Faber}, S.~M., {Gebhardt}, K., {et~al.} 2005, \aj, 129, 2138

\bibitem[{{Lauer} {et~al.}(2007){Lauer}, {Gebhardt}, {Faber}, {Richstone},
  {Tremaine}, {Kormendy}, {Aller}, {Bender}, {Dressler}, {Filippenko}, {Green},
  \& {Ho}}]{Lauer+07a}
{Lauer}, T.~R., {Gebhardt}, K., {Faber}, S.~M., {et~al.} 2007, \apj, 664, 226

\bibitem[{{Law} {et~al.}(2009){Law}, {Kulkarni}, {Dekany}, {Ofek}, {Quimby},
  {Nugent}, {Surace}, {Grillmair}, {Bloom}, {Kasliwal}, {Bildsten}, {Brown},
  {Cenko}, {Ciardi}, {Croner}, {Djorgovski}, {van Eyken}, {Filippenko}, {Fox},
  {Gal-Yam}, {Hale}, {Hamam}, {Helou}, {Henning}, {Howell}, {Jacobsen},
  {Laher}, {Mattingly}, {McKenna}, {Pickles}, {Poznanski}, {Rahmer}, {Rau},
  {Rosing}, {Shara}, {Smith}, {Starr}, {Sullivan}, {Velur}, {Walters}, \&
  {Zolkower}}]{Law+09}
{Law}, N.~M., {Kulkarni}, S.~R., {Dekany}, R.~G., {et~al.} 2009, \pasp, 121,
  1395

\bibitem[{{Magorrian} \& {Tremaine}(1999)}]{MagTre99}
{Magorrian}, J., \& {Tremaine}, S. 1999, \mnras, 309, 447

\bibitem[{{Maksym} {et~al.}(2010){Maksym}, {Ulmer}, \& {Eracleous}}]{Maksym+10}
{Maksym}, W.~P., {Ulmer}, M.~P., \& {Eracleous}, M. 2010, \apj, 722, 1035

\bibitem[{{Maksym} {et~al.}(2013){Maksym}, {Ulmer}, {Eracleous}, {Guennou}, \&
  {Ho}}]{Maksym+13}
{Maksym}, W.~P., {Ulmer}, M.~P., {Eracleous}, M.~C., {Guennou}, L., \& {Ho},
  L.~C. 2013, \mnras, 435, 1904

\bibitem[{{McConnell} \& {Ma}(2013)}]{McCMa13}
{McConnell}, N.~J., \& {Ma}, C.-P. 2013, \apj, 764, 184

\bibitem[{{Merritt}(2006)}]{Merrit07}
{Merritt}, D. 2006, \apj, 648, 976

\bibitem[{{Merritt} \& {Poon}(2004)}]{MerPoo04}
{Merritt}, D., \& {Poon}, M.~Y. 2004, \apj, 606, 788

\bibitem[{{Morris} \& {Serabyn}(1996)}]{MorSer96}
{Morris}, M., \& {Serabyn}, E. 1996, \araa, 34, 645

\bibitem[{{Peng} {et~al.}(2002){Peng}, {Ho}, {Impey}, \& {Rix}}]{Peng+02}
{Peng}, C.~Y., {Ho}, L.~C., {Impey}, C.~D., \& {Rix}, H.-W. 2002, \aj, 124, 266

\bibitem[{{Perets} {et~al.}(2007){Perets}, {Hopman}, \&
  {Alexander}}]{Perets+07}
{Perets}, H.~B., {Hopman}, C., \& {Alexander}, T. 2007, \apj, 656, 709

\bibitem[{{Pracy} {et~al.}(2012){Pracy}, {Owers}, {Couch}, {Kuntschner},
  {Bekki}, {Briggs}, {Lah}, \& {Zwaan}}]{Pracy+12}
{Pracy}, M.~B., {Owers}, M.~S., {Couch}, W.~J., {et~al.} 2012, \mnras, 420,
  2232

\bibitem[{{Pracy} {et~al.}(2013){Pracy}, {Croom}, {Sadler}, {Couch},
  {Kuntschner}, {Bekki}, {Owers}, {Zwaan}, {Turner}, \& {Bergmann}}]{Pracy+13}
{Pracy}, M.~B., {Croom}, S., {Sadler}, E., {et~al.} 2013, \mnras, 432, 3131

\bibitem[{{Quintero} {et~al.}(2004){Quintero}, {Hogg}, {Blanton}, {Schlegel},
  {Eisenstein}, {Gunn}, {Brinkmann}, {Fukugita}, {Glazebrook}, \&
  {Goto}}]{Quinte+04}
{Quintero}, A.~D., {Hogg}, D.~W., {Blanton}, M.~R., {et~al.} 2004, \apj, 602,
  190

\bibitem[{{Rees}(1988)}]{Rees88}
{Rees}, M.~J. 1988, \nat, 333, 523

\bibitem[{{Risaliti} {et~al.}(1999){Risaliti}, {Maiolino}, \&
  {Salvati}}]{1999ApJ...522..157R}
{Risaliti}, G., {Maiolino}, R., \& {Salvati}, M. 1999, \apj, 522, 157

\bibitem[{{Snyder} {et~al.}(2011){Snyder}, {Cox}, {Hayward}, {Hernquist}, \&
  {Jonsson}}]{Snyder+11}
{Snyder}, G.~F., {Cox}, T.~J., {Hayward}, C.~C., {Hernquist}, L., \& {Jonsson},
  P. 2011, \apj, 741, 77

\bibitem[{{Stone} \& {Loeb}(2011)}]{StoLoe11}
{Stone}, N., \& {Loeb}, A. 2011, \mnras, 412, 75

\bibitem[{{Stone} \& {Metzger}(2016)}]{StoMet14}
{Stone}, N.~C., \& {Metzger}, B.~D. 2016, \mnras, 455, 859

\bibitem[{{Strubbe} \& {Quataert}(2009)}]{StrQua09}
{Strubbe}, L.~E., \& {Quataert}, E. 2009, \mnras, 400, 2070

\bibitem[{{Swinbank} {et~al.}(2012){Swinbank}, {Balogh}, {Bower}, {Zabludoff},
  {Lucey}, {McGee}, {Miller}, \& {Nichol}}]{Swinba+12}
{Swinbank}, A.~M., {Balogh}, M.~L., {Bower}, R.~G., {et~al.} 2012, \mnras, 420,
  672

\bibitem[{{Syer} \& {Ulmer}(1999)}]{SyeUlm99}
{Syer}, D., \& {Ulmer}, A. 1999, \mnras, 306, 35

\bibitem[{{Thompson} {et~al.}(2005){Thompson}, {Quataert}, \&
  {Murray}}]{Thomps+05}
{Thompson}, T.~A., {Quataert}, E., \& {Murray}, N. 2005, \apj, 630, 167

\bibitem[{{van der Wel} {et~al.}(2012){van der Wel}, {Bell}, {H{\"a}ussler},
  {McGrath}, {Chang}, {Guo}, {McIntosh}, {Rix}, {Barden}, {Cheung}, {Faber},
  {Ferguson}, {Galametz}, {Grogin}, {Hartley}, {Kartaltepe}, {Kocevski},
  {Koekemoer}, {Lotz}, {Mozena}, {Peth}, \& {Peng}}]{2012ApJS..203...24V}
{van der Wel}, A., {Bell}, E.~F., {H{\"a}ussler}, B., {et~al.} 2012, \apjs,
  203, 24

\bibitem[{{van Velzen} \& {Farrar}(2014)}]{vanFar14}
{van Velzen}, S., \& {Farrar}, G.~R. 2014, \apj, 792, 53

\bibitem[{{van Velzen} {et~al.}(2011){van Velzen}, {Farrar}, {Gezari},
  {Morrell}, {Zaritsky}, {{\"O}stman}, {Smith}, {Gelfand}, \&
  {Drake}}]{vanVel+11}
{van Velzen}, S., {Farrar}, G.~R., {Gezari}, S., {et~al.} 2011, \apj, 741, 73

\bibitem[{{Vanden Berk} {et~al.}(2001){Vanden Berk}, {Richards}, {Bauer},
  {Strauss}, {Schneider}, {Heckman}, {York}, {Hall}, {Fan}, {Knapp},
  {Anderson}, {Annis}, {Bahcall}, {Bernardi}, {Briggs}, {Brinkmann}, {Brunner},
  {Burles}, {Carey}, {Castander}, {Connolly}, {Crocker}, {Csabai}, {Doi},
  {Finkbeiner}, {Friedman}, {Frieman}, {Fukugita}, {Gunn}, {Hennessy},
  {Ivezi{\'c}}, {Kent}, {Kunszt}, {Lamb}, {Leger}, {Long}, {Loveday}, {Lupton},
  {Meiksin}, {Merelli}, {Munn}, {Newberg}, {Newcomb}, {Nichol}, {Owen}, {Pier},
  {Pope}, {Rockosi}, {Schlegel}, {Siegmund}, {Smee}, {Snir}, {Stoughton},
  {Stubbs}, {SubbaRao}, {Szalay}, {Szokoly}, {Tremonti}, {Uomoto}, {Waddell},
  {Yanny}, \& {Zheng}}]{VandenBerk2001}
{Vanden Berk}, D.~E., {Richards}, G.~T., {Bauer}, A., {et~al.} 2001, \aj, 122,
  549

\bibitem[{{Vasiliev} \& {Merritt}(2013)}]{VasMer13}
{Vasiliev}, E., \& {Merritt}, D. 2013, \apj, 774, 87

\bibitem[{{Voges} {et~al.}(1999){Voges}, {Aschenbach}, {Boller},
  {Br{\"a}uninger}, {Briel}, {Burkert}, {Dennerl}, {Englhauser}, {Gruber},
  {Haberl}, {Hartner}, {Hasinger}, {K{\"u}rster}, {Pfeffermann}, {Pietsch},
  {Predehl}, {Rosso}, {Schmitt}, {Tr{\"u}mper}, \& {Zimmermann}}]{Voges99}
{Voges}, W., {Aschenbach}, B., {Boller}, T., {et~al.} 1999, \aap, 349, 389

\bibitem[{{Wang} \& {Merritt}(2004)}]{WanMer04}
{Wang}, J., \& {Merritt}, D. 2004, \apj, 600, 149

\bibitem[{{Wegg} \& {Bode}(2011)}]{WegBod11}
{Wegg}, C., \& {Bode}, J. 2011, \apjl, 738, L8

\bibitem[{{Wyder} {et~al.}(2007){Wyder}, {Martin}, {Schiminovich}, {Seibert},
  {Budav{\'a}ri}, {Treyer}, {Barlow}, {Forster}, {Friedman}, {Morrissey},
  {Neff}, {Small}, {Bianchi}, {Donas}, {Heckman}, {Lee}, {Madore}, {Milliard},
  {Rich}, {Szalay}, {Welsh}, \& {Yi}}]{2007ApJS..173..293W}
{Wyder}, T.~K., {Martin}, D.~C., {Schiminovich}, D., {et~al.} 2007, \apjs, 173,
  293

\bibitem[{{Yang} {et~al.}(2006){Yang}, {Tremonti}, {Zabludoff}, \&
  {Zaritsky}}]{Yang+06}
{Yang}, Y., {Tremonti}, C.~A., {Zabludoff}, A.~I., \& {Zaritsky}, D. 2006,
  \apjl, 646, L33

\bibitem[{{Yang} {et~al.}(2004){Yang}, {Zabludoff}, {Zaritsky}, {Lauer}, \&
  {Mihos}}]{Yang+04}
{Yang}, Y., {Zabludoff}, A.~I., {Zaritsky}, D., {Lauer}, T.~R., \& {Mihos},
  J.~C. 2004, \apj, 607, 258

\bibitem[{{York} {et~al.}(2000){York}, {Adelman}, {Anderson}, {Anderson},
  {Annis}, {Bahcall}, {Bakken}, {Barkhouser}, {Bastian}, {Berman}, {Boroski},
  {Bracker}, {Briegel}, {Briggs}, {Brinkmann}, {Brunner}, {Burles}, {Carey},
  {Carr}, {Castander}, {Chen}, {Colestock}, {Connolly}, {Crocker}, {Csabai},
  {Czarapata}, {Davis}, {Doi}, {Dombeck}, {Eisenstein}, {Ellman}, {Elms},
  {Evans}, {Fan}, {Federwitz}, {Fiscelli}, {Friedman}, {Frieman}, {Fukugita},
  {Gillespie}, {Gunn}, {Gurbani}, {de Haas}, {Haldeman}, {Harris}, {Hayes},
  {Heckman}, {Hennessy}, {Hindsley}, {Holm}, {Holmgren}, {Huang}, {Hull},
  {Husby}, {Ichikawa}, {Ichikawa}, {Ivezi{\'c}}, {Kent}, {Kim}, {Kinney},
  {Klaene}, {Kleinman}, {Kleinman}, {Knapp}, {Korienek}, {Kron}, {Kunszt},
  {Lamb}, {Lee}, {Leger}, {Limmongkol}, {Lindenmeyer}, {Long}, {Loomis},
  {Loveday}, {Lucinio}, {Lupton}, {MacKinnon}, {Mannery}, {Mantsch}, {Margon},
  {McGehee}, {McKay}, {Meiksin}, {Merelli}, {Monet}, {Munn}, {Narayanan},
  {Nash}, {Neilsen}, {Neswold}, {Newberg}, {Nichol}, {Nicinski}, {Nonino},
  {Okada}, {Okamura}, {Ostriker}, {Owen}, {Pauls}, {Peoples}, {Peterson},
  {Petravick}, {Pier}, {Pope}, {Pordes}, {Prosapio}, {Rechenmacher}, {Quinn},
  {Richards}, {Richmond}, {Rivetta}, {Rockosi}, {Ruthmansdorfer}, {Sandford},
  {Schlegel}, {Schneider}, {Sekiguchi}, {Sergey}, {Shimasaku}, {Siegmund},
  {Smee}, {Smith}, {Snedden}, {Stone}, {Stoughton}, {Strauss}, {Stubbs},
  {SubbaRao}, {Szalay}, {Szapudi}, {Szokoly}, {Thakar}, {Tremonti}, {Tucker},
  {Uomoto}, {Vanden Berk}, {Vogeley}, {Waddell}, {Wang}, {Watanabe},
  {Weinberg}, {Yanny}, {Yasuda}, \& {SDSS Collaboration}}]{York+00}
{York}, D.~G., {Adelman}, J., {Anderson}, Jr., J.~E., {et~al.} 2000, \aj, 120,
  1579

\end{thebibliography}

\end{document}